\documentclass[%
 reprint,
superscriptaddress,
 amsmath,amssymb,
 aps,
pra,
]{revtex4-1}

\usepackage{setspace}
\usepackage{graphicx}
\usepackage{bbold}
\usepackage{amsmath}
\usepackage{amsfonts}
\usepackage{amsthm}
\usepackage{float}
\usepackage[margin=1in]{geometry}
\usepackage{enumerate}
\usepackage{etoolbox}
\usepackage{multirow}
\usepackage{amssymb}
\usepackage{comment}
\usepackage{mathrsfs}
\usepackage{ gensymb }
\usepackage{braket}
\usepackage{siunitx}
\usepackage{dcolumn}
\begin{document}

\newcommand{\moire}[0]{moir\'e\ }

\title{Three Phase-Grating Moir\'e\  Neutron Interferometer for Large Interferometer Area Applications }

\author{D. Sarenac}
\affiliation{Department of Physics, University of Waterloo, Waterloo, ON, Canada, N2L3G1}
\affiliation{Institute for Quantum Computing, University of Waterloo,  Waterloo, ON, Canada, N2L3G1}
\author{D. A. Pushin}
\affiliation{Department of Physics, University of Waterloo, Waterloo, ON, Canada, N2L3G1}
\affiliation{Institute for Quantum Computing, University of Waterloo,  Waterloo, ON, Canada, N2L3G1}
\author{M. G. Huber}
\affiliation{National Institute of Standards and Technology, Gaithersburg, Maryland 20899, USA}
\author{D. S. Hussey}
\affiliation{National Institute of Standards and Technology, Gaithersburg, Maryland 20899, USA}
\author{H. Miao}
\affiliation{Biophysics and Biochemistry Center, National Heart, Lung and Blood Institute, National Institutes of Health, Bethesda, Maryland USA}
\author{M. Arif}
\affiliation{National Institute of Standards and Technology, Gaithersburg, Maryland 20899, USA}
\author{D. G. Cory}
\affiliation{Institute for Quantum Computing, University of Waterloo,  Waterloo, ON, Canada, N2L3G1} 
\affiliation{Department of Chemistry, University of Waterloo, Waterloo, ON, Canada, N2L3G1}
\affiliation{Perimeter Institute for Theoretical Physics, Waterloo, ON, Canada, N2L2Y5}
\affiliation{Canadian Institute for Advanced Research, Toronto, Ontario, Canada, M5G 1Z8}
\author{A. D. Cronin} 
\affiliation{University of Arizona, Department of Physics, Tucson, AZ, USA, 85721}
\author{B. Heacock}
\affiliation{Department of Physics, North Carolina State University, Raleigh, NC 27695, USA}
\affiliation{Triangle Universities Nuclear Laboratory, Durham, North Carolina 27708, USA}
\author{D. L. Jacobson}
\affiliation{National Institute of Standards and Technology, Gaithersburg, Maryland 20899, USA}
\author{J. M. LaManna}
\affiliation{National Institute of Standards and Technology, Gaithersburg, Maryland 20899, USA}
\author{H. Wen}
\affiliation{Biophysics and Biochemistry Center, National Heart, Lung and Blood Institute, National Institutes of Health, Bethesda, Maryland USA}

\begin{abstract}
We demonstrate a three phase-grating neutron interferometer as a robust candidate for large area interferometry applications and  characterization of materials. This novel far-field \moire technique allows for broad wavelength acceptance and relaxed requirements related to fabrication and alignment, circumventing the main obstacles associated with perfect crystal neutron interferometry. Interference fringes were observed with a total interferometer length of four meters, and the effects of an aluminum 6061 alloy sample on the coherence of the system was examined. Experiments to measure the autocorrelation length of samples and the universal gravitational constant are proposed and discussed.
\end{abstract}

\maketitle

\section{Introduction}

Interferometers employing particle self-interference have proven to be an extremely sensitive measuring tool, allowing for the precise characterization of material properties as well as measurements of fundamental constants \cite{ni_book2ed,Cronin_2009_RMP}. Neutrons in particular are a convenient probe due to their relatively large mass, electric neutrality, and sub nanometer sized wavelengths. 
The earliest neutron interferometer (NI) used wavefront division from a pair of prisms
to realize Fresnel interference effects with up to $0.06$~mm path separations \cite{Maier-Leibnitz1962}. Amplitude
division from Bragg diffraction at crystal planes was later used to make a perfect
crystal NI with Mach-Zhender path separations of centimeters \cite{rauch1974test}. The relatively large path separation along with the macroscopic size of the interferometer contributed to its success in exploring the nature of the neutron and its interactions \cite{Rauch_1975_PhysLetta,Werner_1988_Physica,oam,chameleon,holography,denkmayr2014observation}. However, a usable perfect crystal NI is difficult to fabricate and operating it requires stringent forms of vibration isolation, beam collimation, and alignment \cite{Arif_1994_VibrMonitCont,saggu2016decoupling,pushin2015neutron,shahi2016new,zawisky2010large}. 

Microfabricated periodic structures have also been employed as neutron optical elements to produce quantum interference. This led to a demonstration of a Mach-Zehnder based grating NI  with reflection gratings \cite{Ioffe1985b}, and a three transmission phase-grating Mach-Zehnder NI for cold and very cold neutrons \cite{GRUBER1989363,VanDerZouw2000,Schellhorn1997,Klepp2011}. However, the consequent low intensities inherent in these NIs makes it difficult for these grating interferometers to outperform the perfect crystal neutron interferometer.

\begin{figure*}
\centering\includegraphics[width=\linewidth]{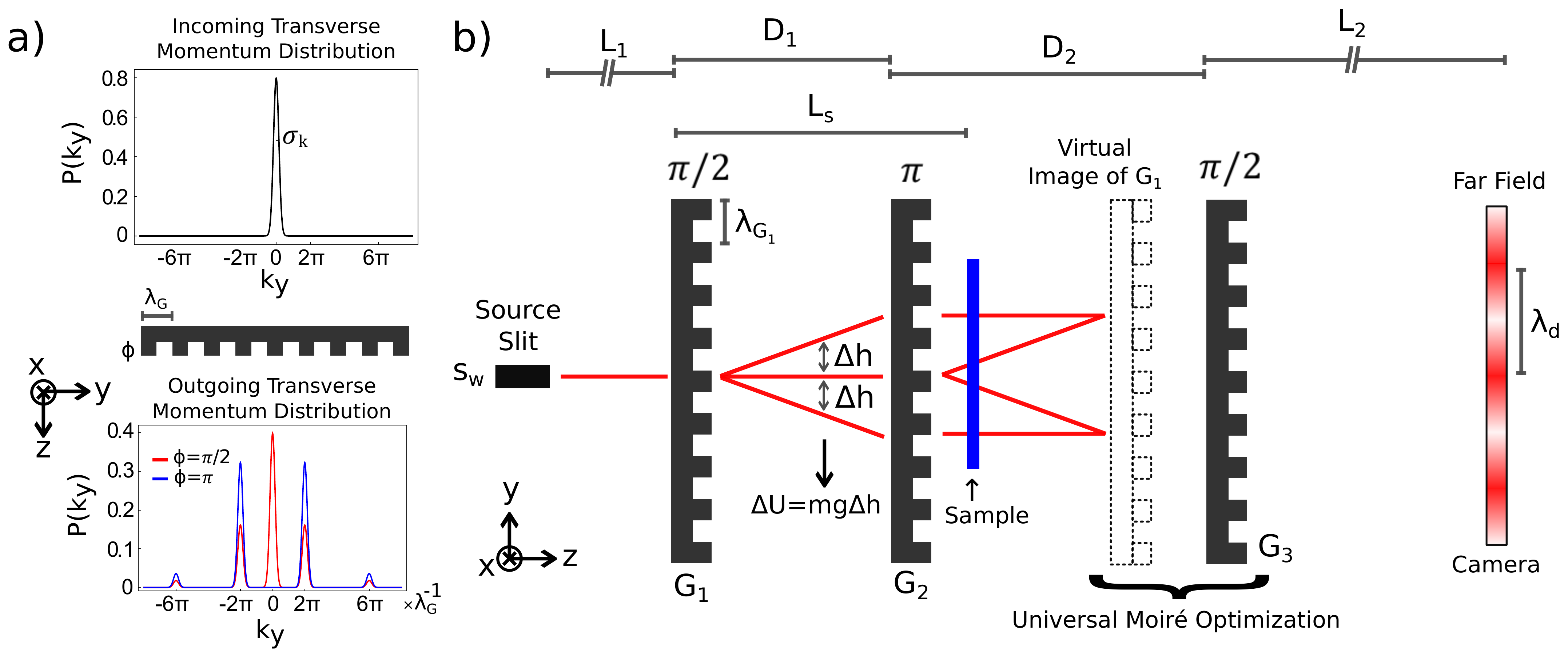}
\caption{a) Writing a phase over the transverse coherence length modifies the neutron's transverse momentum distribution and induces diffraction. Shown is the action of $50$~$\%$ comb-fraction phase-grating whose period ``$\lambda_G$'' is equal to the transverse coherence length ``$\ell_c$'' of the incoming neutron wavepacket, $\lambda_G=\ell_c=1/(2\sigma_k)$. For a $\pi$ phase-grating the $0^{th}$ diffraction order is completely suppressed. b) Three phase-grating interferometer schematic diagram where the $3^{rd}$ grating is offset from the echo-plane to produce the \moire pattern at the camera. The system can be analyzed as the superposition of continuous arrays of Mach-Zehnder interferometers; two of which are illustrated in the figure. This interferometer is sensitive to phase gradients, such as those induced by gravity. A sample may be placed between the gratings for phase imaging, or for dark field imaging.}
 \label{fig:setup}
\end{figure*}

Here we demonstrate a broadband, three phase-grating NI operating in the far field. This NI setup employs the universal \moire effect \cite{miao2016universal} and is an extension to the recently demonstrated two phase-grating \moire interferometer \cite{twogratings,hussey2016demonstration}. Unlike the typical Mach-Zehnder type of interferometers that have two separate and distinct beam paths, these grating far-field interferometers work in the full-field of a cone beam from a finite source, similar to in-line holographic devices. Such full-field systems can be understood intuitively in the framework of Fourier imaging developed by Cowley and Moodie \cite{cowley1957fourier,cowley1960fourier}: the middle grating produces a series of achromatic Fourier images of the first grating at a specific ``echo-plane'' down the beam. The third grating is detuned from the echo-plane to produce a phase \moire effect with the Fourier images, which is observed as a beat pattern in intensity in the far  field. 

The main differences between the current grating far-field interferometer and the neutron Talbot-Lau grating interferometer \cite{Clauser1994,Pfeiffer2006} is that only phase gratings are used, a broader wavelength distribution is accepted, and that the fringes are observed in the far field. The phase \moire effect in the far field produces large period interference fringes that are orders of magnitude larger than the period of the phase modulation, enabling direct detection with an imaging detector without the need for an absorbing analyzer grating. One advantage of an all-phase grating NI is easier grating fabrication especially at smaller grating periods for higher sensitivity. Besides relaxed grating fabrication and alignment, other advantages of this setup include the use of widely available thermal and cold neutron beams, large interferometer area, and broad wavelength acceptance.

\section{Beam coherence and the effects of phase gratings}

The action of the phase-gratings can be understood by noting that writing a phase over the neutron coherence length modifies its momentum. The induced phase shift on a neutron due to a uniform slab of material of thickness D is \cite{ni_book2ed}:
\begin{align}
\phi=Nb_c\lambda D
\end{align}
\noindent where $Nb_c$ is the is the scattering length density of the material (for Silicon $Nb_c\sim2.07\times 10^{14}$~$\si{\per \square \meter}$), and $\lambda$ is the mean neutron wavelength. The momentum operator $\hat{p}=-i\hbar\frac{\partial}{\partial y}$ suggests that a phase ramp induces a momentum shift whereas step gratings correspond to introducing momentum sidebands. Consider a neutron traveling along the z-direction with momentum $\hbar k_z$ incident onto a phase-grating as shown in Fig.~\ref{fig:setup}a. The transverse wavefunction is typically assumed to be a Gaussian: 
\begin{align}
\Psi(y)=\left(\frac{1}{2 \pi \ell_c^2}\right)^{\frac{1}{4}}e^{-\frac{y^2}{4\ell_c^2}}
\label{Eqn:psiz}
\end{align}
\noindent where the coherence length is given by $\ell_c=1/(2\sigma_k)$, and where $\sigma_k$ is the spread of the wavepacket's transverse momentum distribution along the y-direction. The momentum distribution of the outgoing wavepacket is plotted in the lower half of Fig.~\ref{fig:setup}a and it is given by 
	\begin{align}
P(k_y)
    	&= |\mathcal{F}\{\Psi(y) e^{i\frac{\phi}{2} \text{sign}\left[\cos\left(Gy\right)\right]}\}|^2
	\end{align}
\noindent where $\mathcal{F}$ is the Fourier transform, and $G=2\pi/\lambda_G$ is the grating vector where $\lambda_G$ is the grating period. The $1^{st}$ order diffraction peaks are located at:
\begin{align}
k_y=\pm G.
\end{align}

\subsection*{Three phase-grating \moire neutron interferometer}

The interferometer consists of three phase-gratings and a schematic diagram is depicted in Fig.~\ref{fig:setup}b. In relation to the typical MZ interferometer the setup can be viewed as an ``infinite'' array of MZ interferometers. One pair of such virtual MZ interferometers is illustrated, where a ray from the source is diffracted by the $0^{th}$ and $1^{st}$ orders of the gratings into a pair of neutron paths of nearly equal lengths, thus forming a nearly closed loop between the first and the third grating. The angle between the two orders is given by:
\begin{align}
\theta=\sin^{-1}\left(\frac{\lambda}{\lambda_G}\right)
\label{angle}
\end{align}

Each MZ interferometer represents a pair of mutually coherent diffraction pathways through the three gratings at a specific angle from the source. For a polychromatic source, only pathways of nearly identical lengths are mutually coherent. Since phase gratings do not reduce the transmitted flux, flux conservation means that wave interference cannot produce uniform oscillations of intensity over the full field as in the ideal MZ interferometer. Instead, the multitude of interference effects sum to a spatial pattern of intensity when two conditions are met: 1. the system has an appropriate deviation from the perfect symmetry of the MZ interferometer, for example not equidistant between the gratings; 2. observation at an appropriate distance from the third grating. 

\begin{figure}
\centering\includegraphics[width=\linewidth]{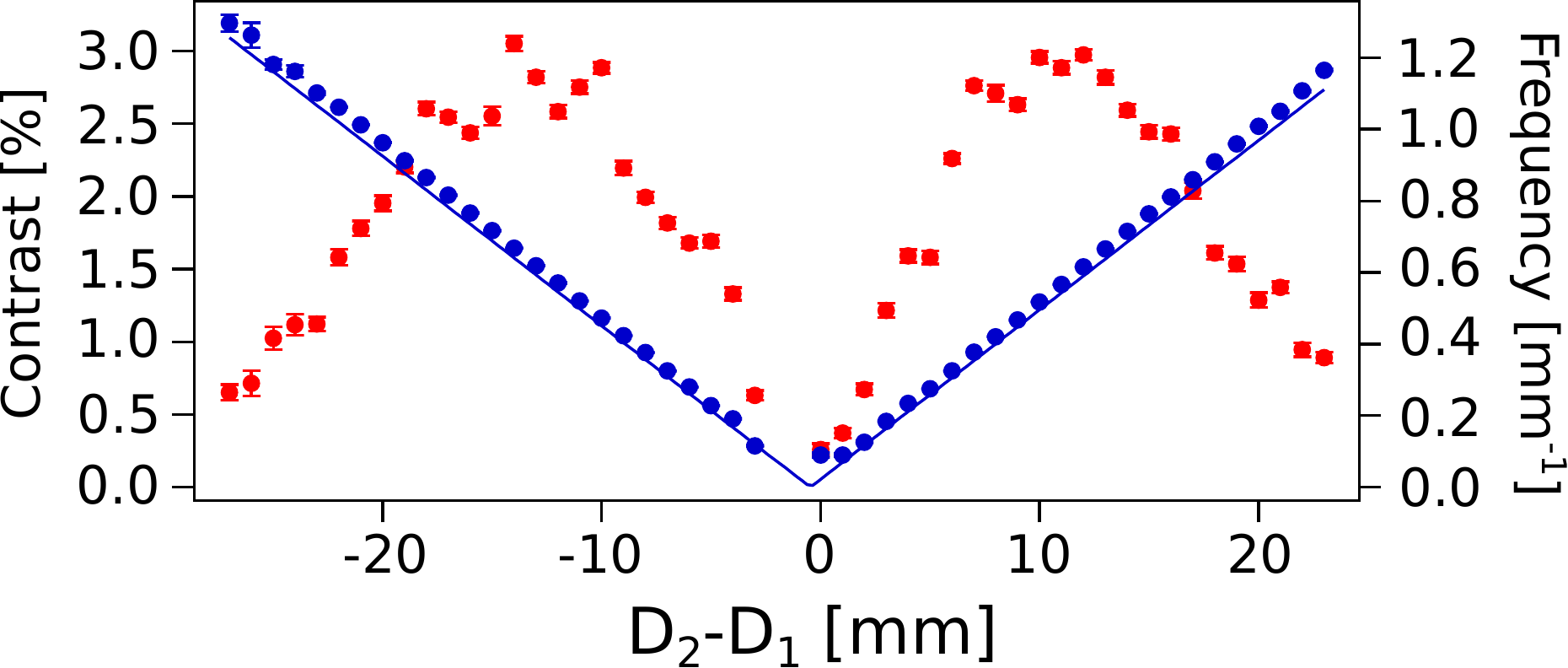}
\caption{The measured contrast (red) and frequency (blue) of the interference pattern at the camera as a function of the difference between the grating separations. The shown uncertainties are purely statistical. The plotted theoretical frequency (straight blue line) derived from Eq.~\ref{Eq:period} shows good agreement with the measured data. When the distance between the $1^{st}$ and $2^{nd}$ grating $(D_1)$ is equal to the distance between the $2^{nd}$ and $3^{rd}$ grating $(D_2)$ the contrast and frequency approach zero. Frequency increases linearly away from that point, while contrast is optimized when $D_2-D_1\approx1.2$~cm for our particular experimental parameters \cite{miao2016universal}. }
 \label{fig:Contrast}
\end{figure}

For highest contrast the first and the third grating should act as $\pi/2$ phase-gratings for the mean wavelength, while the middle grating as a $\pi$ phase-grating. The middle grating acts as a refocusing pulse for the diffracted neutron wavepackets from the first grating. This refocusing, conceptually similar to a spin echo, forms a series of achromatic Fourier images at specific planes downstream \cite{cowley1960fourier}. The third grating then needs to be offset from this Fourier image location in order for the \moire pattern to be observed. When the separation between the first and second grating $D_1$ and the separation between the second and third grating $D_2$ are equal, the image intensity is spatially uniform and flux conservation dictates that the intensity equals the average transmitted intensity through the gratings. As $D_2$ is varied a beating is produced resulting in the \moire pattern at a distance which may be observed by a camera after the third grating.

\begin{figure}
\centering\includegraphics[width=\linewidth]{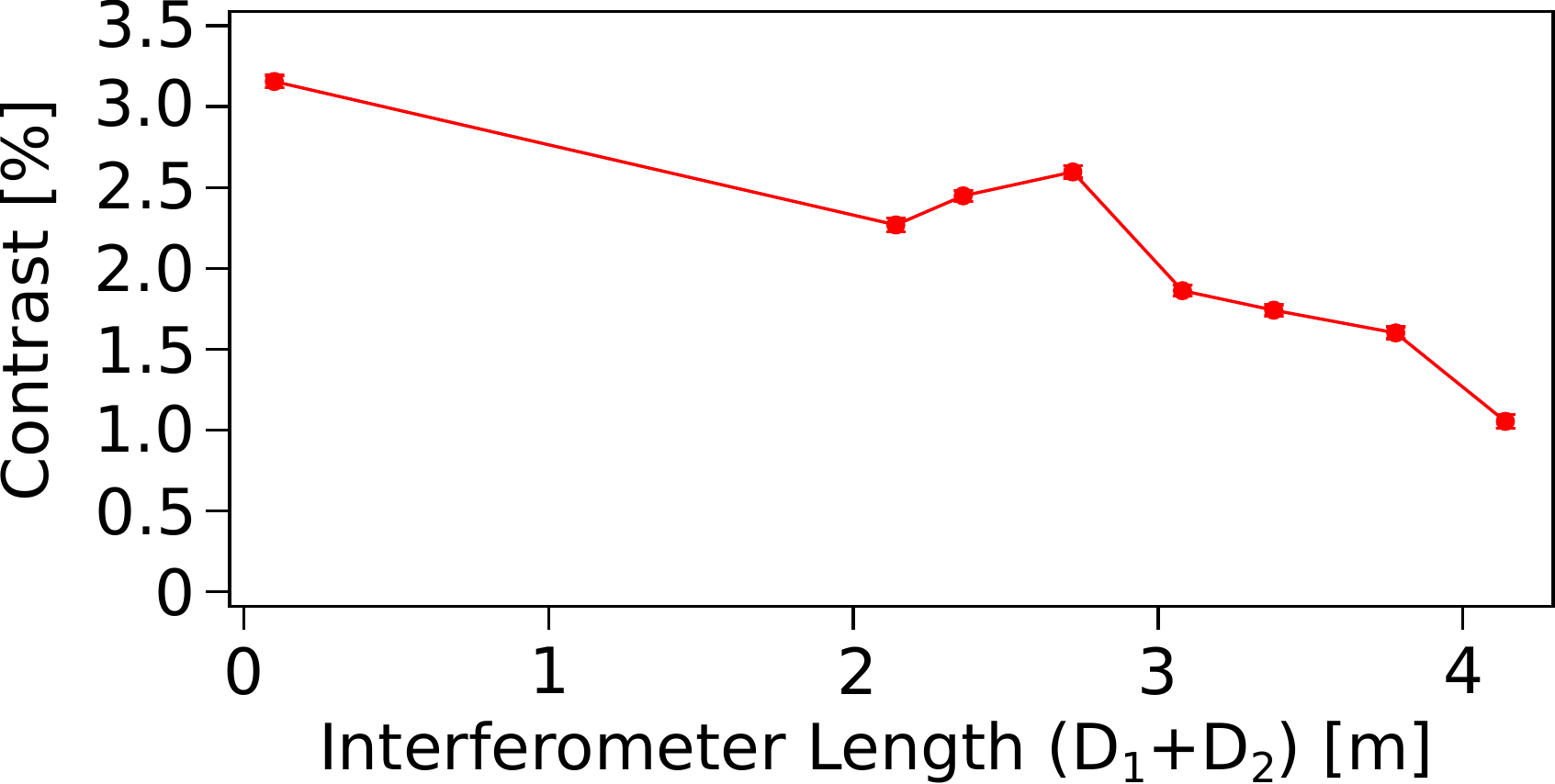}
\caption{Peak contrast, where $D_2-D_1\approx1.2$~cm, as a function of the distance between the $1^{st}$ and $3^{rd}$ grating. The uncertainties are smaller than the individual points and are purely statistical. Interference fringes were observable at a separation of four meters, which was the limit of our experimental setup. This corresponds to an interferometer area of $\sim 8.3$~cm$^2$. Upgrading to a longer beamline allowing for $9$~m interferometer length with $600$~nm phase-grating periods would produce an area of $\sim160$~cm$^2$; an improvement of a factor of $\sim 10$ compare to the standard perfect crystal NI.}
 \label{fig:Contrastvslength}
\end{figure}

An entrance slit defines the transverse coherence length of the neutron wavepackets at the grating location. In order for the neutron wavepacket to diffract, the  coherence length along the grating vector direction (along the y-direction in Fig.~\ref{fig:setup}b) should be at least equal to the period of the grating:
\begin{align}
\ell_c=\frac{\lambda L_1}{s_w} \geqslant \lambda_{\mathrm{G_1}} 
\label{eq:coherence}
\end{align}
\noindent where $L_1$ is the distance form the slit to the first grating, and $s_w$ is the slit width which is the slit opening along the grating vector direction. The slit height $s_h$, which is the slit opening along the perpendicular direction, may be increased in order to increase neutron flux, provided that the gratings are well aligned rotationally with respect to each other.  

The second grating ``$G_2$'', which is ideally a $\pi$ phase grating for the mean wavelength, is placed downstream so that a Fourier image of the first grating is created at a far distance as shown in Fig.~\ref{fig:setup}b. The distance between the two gratings can be substantially varied while maintaining coherence in the system. A third grating ``$G_3$'' is translated around the location of the induced Fourier image to determine the optimal contrast of the fringes at the camera. When all three gratings have the same period the fringe period at the camera is given by \cite{miao2016universal}:
\begin{align}
\lambda_d=\frac{(L_1+D_1+D_2+L_2)}{|D_2-D_1|}\lambda_{G_1}. 
\label{Eq:period}
\end{align}

\noindent The fringe frequency at the detector is given by $f_d=1/\lambda_d$.

\begin{figure}
\centering\includegraphics[width=\linewidth]{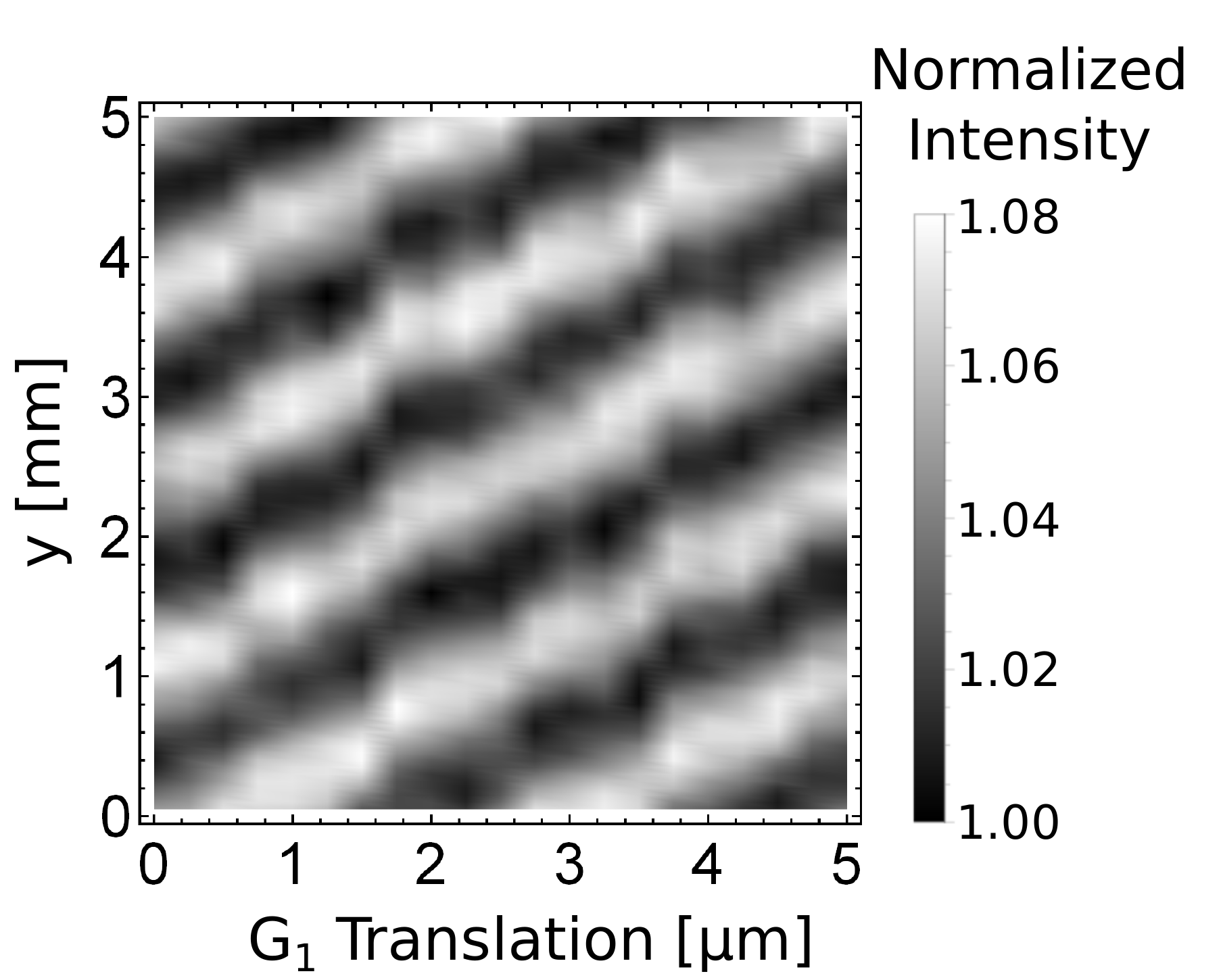}
\caption{Phase stepping. The phase of the interference fringes at the detector is linearly varied by a parallel translation of the $3^{rd}$ grating. The interferometer length was set to $2$~m and $D_2-D_1\approx1.2$~cm to optimize contrast. The $3^{rd}$ grating was then translated along the grating vector (along the y-direction in Fig.~\ref{fig:setup}b) from $0$ to $5$~$\mu$m in increments of $0.25$~$\mu$m. 
}
 \label{fig:PS}
\end{figure}

\section{Experimental Methods}

The experiment was performed at the NG6 Cold Neutron Imaging (CNI) facility \cite{husseybeamline} at the National Institute of Standards and Technology's Center for Neutron Research (NCNR). The neutron spectrum at the CNI is approximately given by a Maxwell-Boltzmann distribution with $T_c=40$~K or $\lambda_c=0.5$~nm. 

We used Si gratings which were available, but not necessarily optimal. The period of each grating was $2.4$~$\mu$m. The $1^{st}$ and $3^{rd}$ grating had a depth of $16$~$\mu$m corresponding to a phase shift of $\sim\pi/2$ for the mean wavelength of $0.5$~nm, while the $2^{nd}$ grating had a depth of $30$~$\mu$m corresponding to a phase shift of $\sim\pi$ for the mean wavelength of $0.5$~nm. The gratings were vertically aligned (with $0.01\degree$ accuracy) to avoid the effects of beam deviation due to gravity. The Si substrate thickness was 550~$\mu$m; the transmittance of neutrons with $\lambda=0.5$~nm through 550~$\mu$m of Si is $\sim99.3\%$.

The slit width was set to $500$~$\mu$m and slit height to $1.5$~cm. The slit to detector length was fixed at $L=8.8$~m, while the distance between the slit and the $2^{nd}$ grating was fixed at $4.75$~m. For Fig.~\ref{fig:Contrast} the distance between the $1^{st}$ and $2^{nd}$ grating was $D_1 = 4.6$~cm, while the distance between $2^{nd}$ and $3^{rd}$ grating, $D_2$, was scanned. For Fig.~\ref{fig:Contrastvslength} the $1^{st}$ and $3^{rd}$ gratings were translated outwards in synchronization to larger separations, and then the $3^{rd}$ grating location was finely scanned. For Fig.~\ref{fig:PS}, we set $D_1= 1$~m and $D_2-D_1\approx1.2$~cm, and then the $3^{rd}$ grating was translated along the grating vector. For Fig.~\ref{fig:sample}, the interferometer length was set to $2$~m.

The imaging detector used was an Andor sCMOS NEO camera viewing a $150~\mu$m thick LiF:ZnS scintillator with a Nikon 50 mm lens, yielding a spatial resolution of  $150~\mu$m \footnote{Certain trade names and company products are mentioned in the text or identified in an illustration in order to adequately specify the experimental procedure and equipment used.  In no case does such identification imply recommendation or endorsement by the National Institute of Standards and Technology, nor does it imply that the products are necessarily the best available for the purpose.}. To reduce noise in the sCMOS system, the median of three images were used for analysis. The exposure time was $20$~s per image, and the detector efficiency was $\eta=0.4$. 

\section{Results and Discussion}

The integrated (along x-direction in Fig.~\ref{fig:setup}b) intensity profile recorded by the camera can be fit to a cosine function

\begin{align}
I=A+B\cos(f_dy+\phi).
\label{eq:sine}
\end{align}

\noindent where $y$ is the pixel location on the camera, $A$ is the mean, $B$ is the amplitude, $f_d$ is the frequency, and $\phi$ is the differential phase. The {\it contrast} or {\it fringe visibility} is given by:
\begin{align}
\mathcal{C}=\frac{\max\{I\}-\min\{I\}}{\max\{I\}+\min\{I\}}=\frac{B}{A}. 
\label{eq:contrast}
\end{align}

In \cite{miao2016universal} it is shown that the contrast is dependent on an autocorrelation of the profiles of $G_1$ and $G_3$, which peaks when the autocorrelation distance is half the grating period for ideal square gratings. For our geometry, $D_2-D_1\sim\pm1.2$~cm values produce autocorrelation distances close to half a period for both $G_1$ and $G_3$. The observed contrast depicted on Fig.~\ref{fig:Contrast} is in agreement with the theory. On Fig.~\ref{fig:Contrast} it can also be seen that the frequency is linearly proportional to the difference of separation distances. At equal separation distances, $D_1=D_2$, the third grating is at the Fourier image location and no fringes are expected.

\begin{figure*}
\centering\includegraphics[width=\linewidth]{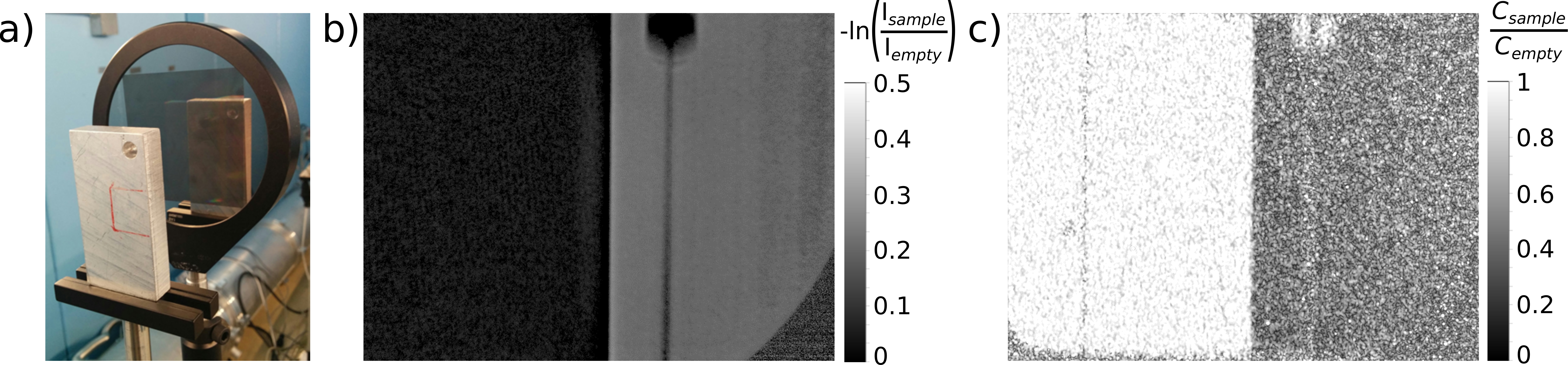}
\caption{Sample imaging. a) A rectangular sample of 6061 Aluminum alloy was placed after the $2^{nd}$ grating. b) Linear attenuation of average intensity calculated as $-\text{ln}(I_{sample}/I_{empty})$. The shape of the sample and the hole in the corner are recognizable in the image. c) Normalized contrast calculated as $\mathcal{C}_{sample}/\mathcal{C}_{empty}$. It can be seen that the alloy destroys the coherence of the system; most likely due to the microstructure of the alloy.}
 \label{fig:sample}
\end{figure*}

The first and third grating can be translated in synchronization from the middle grating in order to achieve large interferometer area. Fig.~\ref{fig:Contrastvslength} shows the peak contrast as a function of the distance from the first to the third grating, which we term as the interferometer length. For this test the middle grating was fixed while the first and third gratings were moved further away from $G_2$ in equally large steps; where at each step the contrast optimization was performed by finely translating the third grating. Contrast was observed with a peak interferometer length of $4$~m. This was the  maximum spacing that could be achieved at CNI. Using Eq.~\ref{angle} an interferometer area of $\sim 8.3$~cm$^2$ is estimated. 

Varying the phase shift of the induced interference fringes is obtained by parallel translation of the third grating with step sizes smaller than the period of the gratings. This so-called ``linear phase stepping'' is shown in Fig.~\ref{fig:PS}. It can be seen that the phase of the interference fringes increases linearly with the grating translation.

Placing a sample between the gratings allows for phase imaging \cite{twogratings,pushin2007reciprocal}. For a rectangular sample of 6061 Aluminum alloy the linear attenuation of average intensity calculated as $\text{ln}(I_{sample}/I_{empty})$, and the normalized contrast calculated as $\mathcal{C}_{sample}/\mathcal{C}_{empty}$, are shown in Fig~\ref{fig:sample}. It is observed that the sample degrades the coherence in the system, most likely due to the microstructure of the alloy. The effects of microstructure are discussed in more detail in the next section. The images were obtained by the Fourier transform method described in \cite{wen2008spatial}.

\subsection*{Potential applications}

The three phase-grating interferometer presents a unique opportunity for material characterization as one can readily vary by orders of magnitude the autocorrelation length used to probe the sample. Although the experiment would be analogous to the probing of autocorrelation lengths with the two phase-grating interferometer \cite{hussey2016demonstration}, the probed autocorrelation length in this case is the separation length of the individual MZ interferometers depicted in Fig.\ref{fig:setup}b:

\begin{align}
\Delta h\sim\frac{\lambda }{\lambda_G}L_s
\label{eq:correlation}
\end{align}
\noindent where $L_s$ is the distance from the first grating to the sample. Therefore the unique ability of the three phase-grating interferometer is accessing larger autocorrelation lengths $(>100$~$\mu m)$, which are beyond the standard limits of the ultra small angle neutron scattering (USANS) and other neutron dark-field imaging methods.

The enclosed area by the interfering neutron paths is an important parameter of a NI and its response to potential gradients and forces. The three phase-grating \moire interferometer has the unique opportunity to reach and surpass the perfect crystal neutron interferometer in this regard. In the current setup, with $2$~m separation between the gratings, the enclosed area is $\approx 8$~cm$^2$ for $0.5$~nm wavelength neutrons, while it is $\approx 15$~cm$^2$ for largest perfect crystal interferometer available at NIST for 0.271~nm neutrons. Reducing the grating period to $600$~nm and  upgrading to a longer beamline which can accommodate grating separation of $4.5$~m will potentially increase the area to $\sim160$~cm$^2$. Another key advantage of the three phase-grating \moire interferometer is in terms of the accepted neutron flux, as the uncertainties in the NI contrast measurements are purely statistical. The neutron acceptance in the perfect crystal setup due to Bragg diffraction is orders of magnitude smaller than the broadband acceptance of the three phase-grating \moire setup.

One of the hallmark neutron interferometer experiment was the ``COW'' experiment (named for the authors  of the first paper: Collella, Overhauser, and Werner) which measured the phase shift of neutrons caused by their interaction with Earth's gravitational field \cite{colella1975observation}, which is a measure of the  standard acceleration due to gravity ``g''. The interferometer used had an area of around $8$~cm$^2$, and the most sensitive versions of the experiment were completed with $\delta g/g \approx 10^{-2}$ agreement with theory, and statistical uncertainty of $\delta g/g \approx 10^{-3}$ \cite{werner1988neutron}.  Recently some of the authors demonstrated that this disagreement may have been due to Bragg-plane misalignments in the interferometer blades \cite{heacock2017neutron}.  Since the original COW experiments, $g$ has been measured using neutrons with a very cold neutron (VCN) interferometer at the $8 \times 10^{-4}$ level \cite{van2000aharonov} and a spin-echo spectrometer at the $10^{-3}$ level \cite{de2014measurement}.

The three phase-grating \moire NI allows for a similar experiment, where the gratings are rotated in synchronization around the beam axis as to vary the angle of the diffracted path, and thereby the induced gravitational potential. Considering only the current neutron beam setup with neutron fluence rate at the slit $ 10^7 \, \si{\per \square \milli \meter \per \second}$ and the slit $15$~mm by $0.5$~mm will yield an incoming flux of $N\approx7.5\times10^7$~$\si{\per  \second}$. With current contrast $C=0.01$ and detector efficiency $\eta=0.4$, the uncertainty $\delta \phi$ in the phase ($\phi$) due to counting statistics (shot noise) is:
\begin{align}
\delta \phi=\frac{1}{C\sqrt{\eta N t}}\approx 2.4\times 10^{-3}\, \si{\radian}
\label{eq:PhaseError}
\end{align}
\noindent in a $t=1$ minute measurement time. The phase due to Earth's gravitational acceleration ($g=9.8 \,\si{\meter \per \square \second}$) is:
\begin{align}
\phi=g T^2 k_g\approx 160\, \si{\radian},
\label{eq:PhaseEarth}
\end{align}
\noindent where $k_g=2\pi/\lambda_G$ with grating period $\lambda_G=2.4 \, \si{\micro \meter}$,  and $T=D_{12}/v_n$ is the neutron flight time between the gratings where $v_n\approx800\, \si{\meter \per \second}$ is the peak neutron velocity.
Thus one minute of measurement in the current setup would offer:
\begin{align}
\frac{\delta \phi}{\phi}=\frac{\delta g}{g} \approx 1.5\times10^{-5}.
\label{eq:PhaseEarth}
\end{align}

There are many aspects of the interferometer that we can improve and expand. These include interferometer contrast (theoretically possible $32$~$\%$), grating period ($\SI{600}{\nano \meter}$ period is currently achievable), and interferometer length up to $\SI{9}{\meter}$ total length. Each of these will improve $\delta g/g$ sensitivity.

Furthermore, a successful realization of the COW experiment could lead to a similar experiment to measure big ``G'', the Newtonian constant of gravitation. The CODATA recommended value of G$=6.67408(31)\times 10^{-11}$~$\mathrm{m}^3 \mathrm{kg}^{-1} \mathrm{s}^{-2}$ with relative standard uncertainty of $4.7 \times 10^{-5}$ \cite{CODATA2014} consists of several discrepant experimental results. We can take advantage of the long path of the interferometer to place a large mass along the neutron paths. In principle this would allow for a measurement using the three phase-grating NI of $\delta G/G$ to a $10^{-5}$ level or smaller.

\section{Conclusion}

For the first time we have demonstrated a \moire effect neutron interferometer consisting of three phase-gratings. The interferometer has a broad wavelength acceptance, requires non-rigorous alignment, and operates in the far-field regime. These advantages make it possible to circumvent many limitations of the contemporary single crystal Mach-Zehnder type, and the near field Talbot-Lau type neutron interferometers. 

Although this initial demonstration achieved a maximum $3$~$\%$ contrast, the theoretical maximum contrast for square profile gratings is $32$~$\%$ \cite{miao2016universal}. The factors that reduce the contrast are the finite slit width, which is estimated at a relative fraction of between $11$~$\%$ and $18$~$\%$ depending on the slit transmission profile; the actual phase-shift profile of $G_2$ which determines its efficiency; neutron scattering over the long distance of the NI by air or intervening parts such as vacuum windows. Contrast reduction with increased grating separation also points to scattering effects over distance and mechanical vibration as potential factors that degrade performance \cite{bushuev2016dynamic}. Future work will include direct assessment of individual grating diffraction efficiencies to characterize and minimize these losses. 

We expect that the next generation of interferometers based on the three phase-grating far-field design will open new opportunities for the characterization of materials with a large autocorrelation function, and for the measuring of the fundamental gravitational constants and other small forces. 

\section{Acknowledgments}

This work was supported by the U.S. Department of Commerce, the NIST Radiation and Physics Division, the Director's office of NIST, the NIST Center for Neutron Research, and the National Institute of Standards and Technology (NIST) Quantum Information Program. This work was also supported by the Canadian Excellence Research Chairs (CERC) program, the Canada  First  Research  Excellence  Fund  (CFREF), the Natural Sciences and Engineering Research Council of Canada (NSERC) Discovery program, US Department of Energy (DOE), and the Collaborative Research and Training Experience (CREATE) program.

\bibliography{lib}

\end{document}